\documentstyle[psfig]{mn}
\input{psfig.sty}

\newif\ifAMStwofonts

\title[On the interpretation of the multicolour model]
        {On the interpretation of the multicolour disc model for black hole candidates}
\author[A. Merloni et al]
        { A.Merloni$^{1}$,  A.C. Fabian$^{1}$ and
         R.R. Ross$^{2}$
\\$1$ Institute of Astronomy, Madingley Road, Cambridge, CB3 0HA
\\$2$ Physics Department, College of the Holy Cross, Worcester, MA 01610, USA
}

\date{}

\begin{document}

\maketitle

\label{firstpage}

\begin{abstract}
We present a critical analysis of the usual interpretation of the {\it multicolour disc} 
model parameters for black hole candidates in terms of the inner radius and temperature 
of the accretion disc. Using a self-consistent model for the radiative transfer and 
%%vertical structure
the vertical temperature structure in a Shakura-Sunyaev disc, we simulate the observed disc 
spectra, taking into account doppler blurring and gravitational redshift, and 
fit them with multicolour models. We show not only that such a model systematically 
underestimates the value of the inner disc radius, but that when the accretion rate 
and/or the energy dissipated in the corona are allowed to change the inner edge of the 
disc, as inferred from the multicolour model, appears to move even when it is in fact fixed 
at the innermost stable orbit.        
\end{abstract}

\begin{keywords}
accretion, accretion discs - black hole physics - radiative transfer - X-rays: stars
\end{keywords}

\section{Introduction}

X-ray spectra of galactic black hole candidates (GBHC) are becoming an increasingly 
important tool in determining both the physical properties of the source surroundings 
and, as the ultimate goal, the  parameters (mass and specific angular momentum) of the 
black hole. Two main spectral states, as defined by their spectral components and flux 
level (typically in the $1-10$ keV band), have been observed in GBHC. In the 
{\it Hard/Low state}, the sources emit most of their power in a hard power law tail with 
photon index $\Gamma \sim 1.3 - 1.7$ and exponential cutoff at about $100$ keV. In the 
generally more luminous {\it Soft/High state}, most of the energy comes from a thermal 
component with characteristic temperature $kT \le 1$ keV, while a power law component 
(with $\Gamma \sim 2.0 - 2.5$) dominates above a few keV. In addition, sources have been 
reported to be in a {\it Very high state}, spectrally intermediate between the Soft/High 
and the Hard/Low states, when the power law component has a flux comparable to the thermal 
one and no sign of a cutoff is seen in the high energy tail.   

We will focus our discussion on the Soft/High state. According to the common paradigm in 
interpreting the observed spectra, the ultra-soft thermal component is the product of the 
emission from an optically thick accretion disk. On the other hand, the hard power law 
component is probably due to multiple Compton scattering of soft photons by a population 
of hot electrons which reside in an active (or flaring) coronal region surrounding the 
accretion disc. 

%%The zeroth order approximation for describing emission from the accretion disc is 
%%the assumption that the disc is geometrically thin and optically thick \cite{SS73}: 
In the zeroth order approximation for describing emission from the accretion disc, not 
only is the disc assumed to be geometrically thin and optically thick \cite{SS73}, but also 
the vertical temperature structure is neglected, as is the effect of Comptonization on 
the emergent spectrum. Each point of the disc is then assumed to radiate like a blackbody at 
an effective temperature which scales with the radius as $r^{-3/4}$. This is the so-called 
{\it multicolour disc model} (MCD; Mitsuda et al. 1984). Such a simple model has the advantage 
of being easy to use in trying to fit spectral data. It has only two adjustable parameters: 
a ({\it colour}) temperature, independent on the source distance, and a normalization 
factor, which in turn depends on the inner radius of the disc, on the distance of the 
source and on the inclination of the disc from the line of sight. When 
these two latter quantities are independently known, or can be reasonably inferred, fitting 
the ultra-soft component of a GBHC spectrum with the MCD model immediately gives  
estimates of the temperature in the inner region of the disc and its inner radius.

The aim of this letter is to demonstrate that the values one can obtain with this fitting 
procedure, which is now a standard one, are {\it not} in general directly related (or at 
least not in a reliable way) to the actual disc parameters. We will use a self-consistent 
model for the radiative transfer in Shakura-Sunyaev accretion discs about a compact source, 
including first-order corrections for gravitational redshift and Doppler blurring, as 
developed by Ross \& Fabian  \shortcite{RF96}. The observable spectrum is computed for 
%%many 
a number of different physical situations, each determined by the values of the viscosity 
parameter $\alpha$, the accretion rate and the fraction of the accreted power dissipated in 
the corona \cite{SZ94}. In all the cases we consider, the value of the inner radius of the 
disc will be kept fixed. Nevertheless, when trying to fit the computed spectra with the MCD 
model, 
%%as we will show in Section 4, 
we will obtain results at variance with this assumption. That is because changes in the 
normalization factor of the MCD model are produced, in a very complicated way, by the 
variations in the disc structure (and hence in the Comptonized spectrum) induced by 
variability of the accretion rate and coronal activity.

\section[]{Structure of the disc}
\label{struc}

We assume the basic structure for spatially thin accretion discs around Schwarzschild black 
holes given by the standard theory of Shakura \& Sunyaev \shortcite{SS73}. 
The disc is assumed to have a {\it fixed} inner boundary at the innermost stable orbit at 
radius $R_{\rm in}=3R_S=6GM/c^2$. We consider both an inner, radiation pressure dominated 
region and an outer, gas pressure dominated region  
%%and in both we assume the opacity to be 
of the disc. In determining the structure of the disc, the opacity is assumed to be 
dominated by Thomson scattering,
\begin{equation}
\kappa \approx \kappa_T = 0.2 (1+X)\; {\rm cm^2 g ^{-1}} ,
\end{equation}
where $X$ is the mass fraction of hydrogen. Following Svensson \& Zdziarski \shortcite{SZ94}, 
we slightly modify the standard set of disc equations, allowing a fraction $f$ of the disc 
accretion power to be dissipated in the corona instead of in the cold disc itself. 
Furthermore, at every given radius $R$, the density is taken to be uniform in the vertical 
direction (which is the correct solution for the disc structure when radiation pressure 
dominates).

Choosing dimensionless parameters
\[
m=\frac{M}{M_{\odot}} , \;\;  \dot m= \frac{\dot M}{\dot M_{\rm Edd}} , \;\; r=\frac{R}{R_S} ,
\]
where $\dot M_{\rm Edd}=3.1 \times 10^{-8} m \; M_{\odot}\;{\rm yr^{-1}}$ is the Eddington 
accretion rate for a disk efficiency $\eta=0.083$, the flux emerging from the surface of the 
disc is given by 
\begin{equation}
F_0=1.2 \times 10^{27} m^{-1} r^{-3} [\dot m J(r)] (1-f) ,
\end{equation}
with $J(r)=(1-\sqrt{3/r})$. The radial structure of the disc is summarized by the values 
of the half-thickness of the disc (in units of the Schwarzschild radius) $h$ and the 
uniform gas density $\rho_0$:
\begin{enumerate}
\item {{\it Radiation pressure dominated region} 
  \begin{eqnarray}
    \label{Prad}
   && h=\frac{H}{R_S}=5.3 (1+X) [\dot m J(r)] (1-f) \\
   && \rho_0=3.4 \times 10^{-7} \alpha^{-1} m^{-1} r^{3/2} \nonumber \\
 && \times  [\dot m J(r)]^{-2} (1-f)^{-3} {\rm g}\; {\rm cm ^{-3}},
  \end{eqnarray}}
\item {{\it Gas pressure dominated region}
    \begin{eqnarray}
      \label{Pgas}
      && h=3.4 \times 10^{-2} (1+X)^{1/10} \alpha^{-1/10} m^{-1/10} r^{21/20} \nonumber \\ && \times  [\dot m J(r)]^{1/5} (1-f)^{1/10}\\
      && \rho_0=4 (1+X)^{-3/10} \alpha^{-7/10} m^{-7/10} r^{-33/20} \nonumber \\ && \times [\dot m J(r)]^{2/5} (1-f)^{-3/10} {\rm g} \;{\rm cm ^{-3}} .
    \end{eqnarray}}

\end{enumerate}

The radiative transfer and the vertical temperature structure are treated self-consistently 
at each radius, as described in 
%%Ross, Fabian \& Mineshige \shortcite{RFM92}
Ross \& Fabian  \shortcite{RF96}, using the Fokker-Planck/diffusion equation of Ross, 
Weaver \& McCray \shortcite{RWM78} in plane-parallel geometry.
The local temperature profile, $T(z)$, is found by balancing the heating rate due to dynamic 
heating, Compton scattering and free-free absorption with the cooling rate due to inverse 
Compton scattering and free-free emission.
 
%%I'VE REWORKED THE NEXT TWO PARAGRAPHS SOME MORE: 
Incoherent Compton scattering within the accretion disc must be treated properly whenever it 
has an important effect on the emergent spectrum. The competition between Compton scattering 
of higher-energy photons and inverse Compton scattering of lower energy photons often results 
in the emergence of a Wien-law hard tail. The {\it effective optical depth} for 
absorption is given by
\begin{equation}
\tau_{*}(\nu)=\sqrt{3\tau_{\rm ff}(\nu)(\tau_{\rm T}+\tau_{\rm ff}(\nu))} ,
\end{equation}
where $\tau_{\rm T}$ is the Thomson depth below the surface, and $\tau_{\rm ff}(\nu)$ is 
the optical depth due to free-free absorption. For high energy photons the 
{\it thermalization depth} (where $\tau_{*}(\nu)=1$) can be reached at very high 
Thomson depths, where the temperature is considerably higher than near the surface. 
Compton scattering in the outer layers of the disc downscatters these photons to lower 
energies and produces a Wien-law tail.   

%We can say that Comptonization should be complete or `saturated' whenever the {\it thermalization depth} (where $\tau_{*}=1$) for high energy photons is greater than the Comptonization depth (where $y=1$).

Decreasing the flux emerging from the disc itself, either by lowering the accretion rate
($\dot m$) or increasing the fraction ($f$) of the accretion power dissipated in the corona,
results in ever larger regions of very high density, both in the outermost portions of the 
disc and, for very low fluxes, in the very innermost portions as well. In those cases 
Comptonization is not complete or `saturated' because 
\begin{equation}
y=\frac{4kT}{m_ec^2}\tau_T^2 < 1 
\end{equation}
at the thermalization depth. In these regions we treat Compton scattering as {\it coherent}, 
dropping the Fokker-Planck term in the radiative transfer equation (see discussion in 
Ross \& Fabian \shortcite{RF96}).
%%this, in turn, will make the emergent spectrum even harder, and, consequently, the black-body approximation worse.

In our calculations we fix the value of the black hole mass $m=10$ and assume a 
composition $X=0.71$. Following Ross \& Fabian \shortcite{RF96} we divide the region $3<r<200$ 
of the disc into 20 annuli that make comparable contributions to the total luminosity. 
For each set of the parameters $\alpha$, $\dot m$ and $f$, we find the radius at which 
radiation pressure equals gas pressure, which is the root of the equation \cite{SS73}
\begin{equation}
\frac{r_{AB}}{J(r_{AB})^{16/21}}\simeq 370 (m \alpha)^{2/21} \dot m^{16/21} (1-f)^{6/7}.
\end{equation}
We use equations (3) and (4) for all the annuli for which $3<r<r_{AB}$, and 
equations (5) and (6) for $r>r_{AB}$.
The typical emergent spectrum from each annulus is calculated and then multiplied by the area 
of the annulus to find the contribution to the spectral luminosity.
The resulting spectra are added together, taking into account blurring due to 
gravitational redshift and transverse Doppler effect using the method described by 
Chen, Halpern \& Filippenko \shortcite{CHF89}.
Finally, in order to allow direct comparison with observations in which a dramatic change 
in the inner radius has been reported (GRS 1915+105, see e.g.\ Belloni et al. 
\shortcite{Bell97b}), we choose an inclination angle $i=70^{\circ}$ and a distance  
$D=12.5 {\rm kpc}$.

\section{The multicolour disc model}

For each set of parameters the emergent disc spectrum is used as a table model to create a 
fake set of data (with RXTE response matrix p012\_LR1\_970804.rsp, integration time 1000 s).
These data are then fitted  with the standard multicolour blackbody model (DISKBB in XSPEC) 
in the $2-20$ keV band. The model assumes that the local emission from the disc is Planckian, 
with a temperature profile $T(r)\propto r^{-3/4}$. Therefore the observed flux from the disc 
is approximated by
\begin{equation}
F_{d}(E)=\frac{8\pi R_{\rm col}^2 \cos{i}}{3D^2}\int_{T_{\rm out}}^{T_{\rm col}}
\left(\frac{T}{T_{\rm col}}\right)^{-11/3}B(E,T)\frac{dT}{T_{\rm col}} ,
\end{equation}
where $B(E,T)$ is the Planck function and $T_{\rm out}$ is the temperature at the outer radius 
of the disc (and it is assumed that $T_{\rm out}\ll T_{\rm col}$). The two fit parameters of 
the model are the colour temperature of the inner accretion disc ($T_{\rm col}$) in keV and 
the normalization factor
\[
n=\frac{R_{\rm col}^2 \cos{i}}{D^2}  , 
\]
where $R_{\rm col}$ is expressed in km and the distance $D$ in units of $10 {\rm kpc}$.

The parameter $R_{\rm col}$ is {\it not} the effective inner disc radius (i.e.\ the radius at 
which the temperature of the disc is the highest).
%%the Comptonization of the emergent spectrum, which is relevant in the inner region of the disc, forms a Wien peak with characteristic temperature $T_{\rm eff}< T_{\rm col}$. If  the ratio $f_{\rm col}\equiv T_{\rm col}/T_{\rm eff}$ (the so called {\it spectral hardening factor}, Shimura \& Takahara \shortcite{ST95}) was constant throughout the disc and did not vary with varying the physical parameters ($\dot m$, $\alpha$, $f$, etc.), we could still use the multicolour disc model to infer the effective radius, and from that the actual value of the inner edge of the disc: it would be sufficient to rescale
The assumption behind using the MCD model is that the Comptonized emergent spectrum can be 
approximated by a diluted blackbody spectrum with a colour temperature ($T_{\rm col}$) higher
than the effective temperature ($T_{\rm eff}$). The ratio $f_{\rm col}=T_{\rm col}/T_{\rm eff}$, 
the {\it spectral hardening factor} \cite{ST95}, is then assumed to be constant throughout the 
disc and for varying physical parameters ($\dot m$, $\alpha$, $f$, etc.). Thus the actual inner 
edge of the disc would be given by
\begin{equation}
\label{rcol}
R_{\rm in}=\eta R_{\rm eff}=\eta  g(i) R_{\rm col} (T_{\rm col}/T_{\rm eff})^2  ,
\end{equation}
where $\eta$ is 
%%WASN'T THIS SAID BACKWARDS?
the ratio of the inner radius of the disc to the radius at which the emissivity actually peaks, 
typically $\eta \simeq 0.6 - 0.7$ \cite{SS73,ST95}. The factor $(T_{\rm col}/T_{\rm eff})^2$ 
comes from the assumed dilution of the blackbody spectrum, while $g(i)$ takes into 
account general relativistic corrections and is of the order $g \simeq 0.7 - 0.8$ \cite{Ebi94}.

In the next section we will show that these assumptions are not justified in general due to 
the differences in Compton processes induced by changes in $\dot m$, $f$, and $\alpha$.

\section{Results}
We have simulated disc spectra in eleven different physical situations represented by the 
ten sets of parameters S1-S11 (see Table \ref{parameters}). The sets are listed in order of 
decreasing value of the radius $r_{\rm AB}$, the boundary between the radiation pressure and
gas pressure dominated regions. Table \ref{parameters} also lists the value of the radial 
coordinate $r_{\rm coh}$ for the boundary of the region(s) where Compton scattering has been 
treated as coherent (see section \ref{struc}). As the power dissipated in the disc decreases 
and the density increases, from the combined effects of changes in $\dot m$ and $f$, 
this boundary moves inwards.  That is, for an increasing portion of the disc Comptonization is 
incomplete and the local spectrum will be somewhat harder. For the S10 model, the density 
in the innermost region of the disc is so high (due to the $J(r)^{-2}$ term) that we  have to 
consider incomplete Compton scattering also in the first inner annuli, where a significant fraction 
of the X-ray power is emitted. Consequently, for this set of parameters, two values of 
$r_{\rm coh}$ are listed. Finally, for the model S11, when the disc is entirely gas pressure dominated, we considered incomplete Compton scattering throughout all the disc (but see next section for a caveat on the applicability of our model to this extreme case).

Figure 1 shows the observed spectra for S1, S4, S7 and S10, taking into account Doppler 
blurring and gravitational redshift. For comparison, Fig.\ 1 also shows the blurred spectra 
that would be observed if each point on the surface of the accretion disc emitted a blackbody 
spectrum at the local effective temperature. The observed spectra are harder than such 
effective blackbody spectra, and the discrepancy increases as we increase the density of the 
disc and decrease the power released in the disc itself (going from S1 to S10).

\begin{figure*}
\vbox to140mm{\vfil 
\psfig{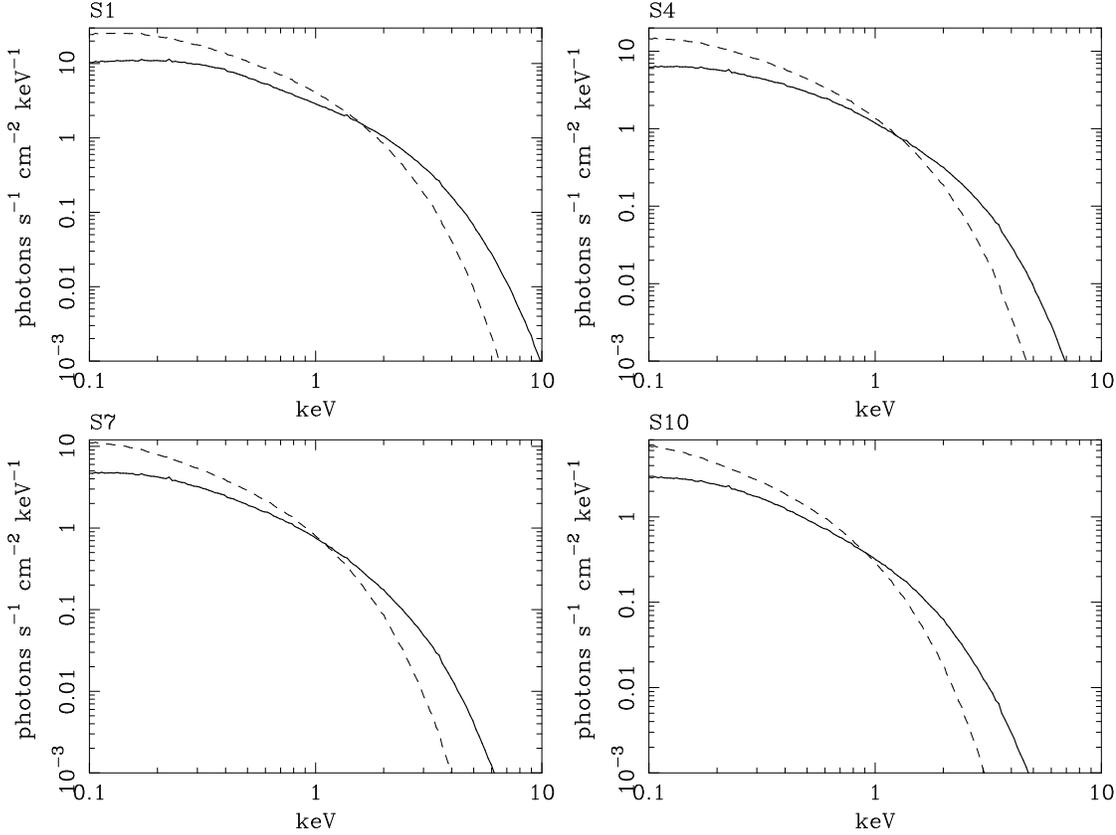}
\caption{The disc spectra (solid lines) computed for the parameter sets S1 ($\dot m=0.3$, $f=0$, $\alpha=0.1$); S4 ($\dot m=0.2$, $f=0.5$, $\alpha=0.3$); S7 ($\dot m=0.1$, $f=0.4$, $\alpha=0.1$) and S10 ($\dot m=0.05$, $f=0.5$, $\alpha=0.1$). For each of the plots also shown are the spectra that would be observed for a blackbody-emitting disc (dashed lines).}
\vfil}
\label{fig1}
\end{figure*}

\begin{table}
 \caption{Physical parameters.}
 \label{parameters}
 \begin{tabular}{@{}lccccc}
      SET   & $\alpha$ & $\dot m$ & $f$ & $r_{\rm AB}$ & $r_{\rm coh}$ \\ 
  S1    & 0.1 & 0.3 & 0.0 & 100 & 50 \\ 
  S2    & 0.1 & 0.2 & 0.0 & 70  & 36 \\
  S3    & 0.1 & 0.1 & 0.0 & 50  & 28 \\
  S4    & 0.1 & 0.2 & 0.5 & 50  & 18 \\ 
  S5    & 0.3  & 0.08 & 0.1 & 36 & 22 \\
  S6    & 0.3  & 0.08 & 0.2 & 36 & 18 \\
  S7    & 0.1  & 0.1 & 0.4 & 28  & 13 \\
  S8    & 0.1  & 0.05 & 0.0 & 28 & 13\\
  S9    & 0.1  & 0.1 & 0.5 & 22  & 9.6\\
  S10   & 0.1  & 0.05  & 0.5 & 13 & 4.6 - 5.2  \\
  S11   & 0.1  & 0.05  & 0.8 & -  & -   \\	
 \end{tabular}

 \medskip
 
\end{table}

\begin{table}
 \caption{Fit parameters in the range 2-20 keV. All the models refer to a $10M_{\odot}$ black hole with inclination angle and distance set to the values of the black hole candidate GRS 1915+105, namely $i = 70^{\circ}$ and $D = 12.5 {\rm kpc}$.}
 \label{results1}
 \begin{tabular}{@{}lcccc}
      SET   & $T_{\rm col} [{\rm keV}]$
        & $R_{\rm col} [{\rm km}]$
        & $\chi^2_{\rm red}$ & $f_{\rm col}$ 
        \\ 
  S1    & 0.98 & 54.5 & 47 & 1.80 \\ 
  S2    & 0.87 & 56.9 & 22 & 1.76 \\
  S3    & 0.75 & 53.3 & 6.0 & 1.82 \\
  S4    & 0.75 & 52.1 & 6.4 & 1.84 \\
  S5    & 0.70 & 49.2 & 2.7 & 1.90 \\
  S6    & 0.69 & 49.3 & 2.5 & 1.90 \\
  S7    & 0.69 & 46.5 & 2.8 & 1.95 \\
  S8    & 0.68 & 41.1 & 1.6 & 2.08 \\
  S9    & 0.67 & 41.5 & 1.5 & 2.07 \\
  S10   & 0.59 & 39.0 & 1.8 & 2.12 \\
  S11   & 0.55 & 24.5 & 0.7 & 2.68 \\
 \end{tabular}

 \medskip
 
\end{table}

Table \ref{results1} lists the values of the fit parameters $T_{\rm col}$ and $R_{\rm col}$ 
for the ten cases, as well as the corresponding reduced $\chi^2$ of the fits and the spectral 
hardening factors obtained from eq. (\ref{rcol}) by assuming $\eta g(70^{\circ})=0.5$ and 
$R_{\rm in}= 88.6 {\rm km}$, 
as appropriate for the $10 M_{\odot}$ Schwarzschild black hole we are considering. 
The results are also shown in Figures 2a and 2b, where we plot the spectral hardening factor 
and the colour radius as functions of $\dot m (1-f)$, which is a measure of the flux emergent 
from the disc itself.

%%We would also like to stress that, 
To check the sensitivity of our results to the energy range 
fitted, we also fitted the spectra in the $3-10$ keV band. The values obtained for the colour 
radii were, at most, just  one or two kilometers smaller than those obtained in the 2-20 keV 
band (i.e.\ a change of about $3-4 \%$).

\begin{figure}
\vbox to140mm {\vfil 
\psfig{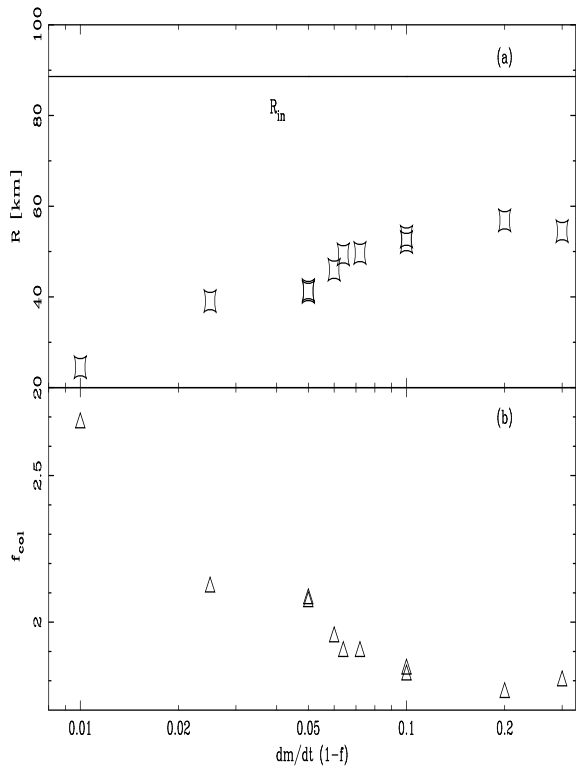}
\caption{(a): the points are the values or $R_{\rm col}$ (in km) plotted as functions of $\dot m (1-f)$, which is a measure of the flux emergent from the disc, while the solid line marks the actual inner boundary of the disc, fixed in all our ten simulations. (b): triangles represent the value of the spectral hardening factor needed to rescale the computed $R_{\rm col}$ to $R_{\rm in}$.}
\vfil}
\label{fig1}
\end{figure}

It is clear that the multicolour model gives systematically lower values for the ``inner radius'' 
than the actual ones, a point which is well known. More interestingly, our results show that 
the hardening factor is by no means constant when we change the physical parameters of the 
disc. For example, assuming the conventional value $f_{\rm col}=1.7 \pm 0.2$ \cite{ST95,Sob99} 
and taking Doppler blurring and gravitational redshift into account, we would have 
underestimated the value for the inner disc radius by up to a factor of $2$ (S11) using the 
fits with a multicolour model. On the other hand the model gives quite stable and 
acceptable results for high accretion rates and/or lower values of the fraction of the power 
eventually dissipated in the corona. This in turn can help to understand the cases in which 
a nearly constant value of $R_{\rm col}$ is observed even when $T_{\rm col}$ varies \cite{TL95}. 

\section{Discussion}

We have shown that MCD models systematically underestimate the value of the inner radius of 
the accretion disc for a black hole candidate. We also have shown that the spectral hardening 
factor $f_{\rm col}$, which is needed to correct the results of the fits with a multicolour 
model, is not, as 
is usually assumed, constant when the accretion rate and/or fractional coronal activity change. 

Recent observations of galactic black hole candidates seem to point towards an extreme dependence of the observed {\it colour} radius on $f$. 

 In the work of Sobczak et al. \shortcite{Sob99}, where the authors report on the RXTE spectral observations of the 1996-97 
outbursts of the microquasar GRO J1655-40, a sudden jump inward of the color radius (see their Fig. 7) occur  for the five very high state observations of the source during which the power-law (hard) flux was extremely high and the blackbody--to--total flux ratio was less than $0.5$ (see their Fig. 5).

A similar trend has been reported by Muno et al. \shortcite{mun99} in the case of the microquasar GRS 1915+105. In their observations, the smallest values of the inner disc radius obtained by fitting the spectrum with  the MCD model are associated with the highest values of the power-law--to--blackbody flux ratio, which in turn should be directly related to the value of $f$.

In both these cases the inferred inner disc radii can change by more than a factor of four. 

Belloni et al. (1997a,1997b), analyzing RXTE data of GRS 1915+105, interpreted the sudden change measured in the inner colour radius of roughly a factor of four in terms of the rapid disappearence of the whole inner part of the geometrically thin accretion disc. Once again, in that case the inner edge of the disc appears to shrink to its smallest values during an outburst in which the power law flux dominates the blackbody, and the blackbody--to--total flux ratio is less than $0.5$.    

This behaviour is exactly 
what our results predict we should expect from the MCD model when we increase $f$ and/or reduce the accretion rate. In this case, every time we the observations imply that the coronal activity is dominant, the multicolour fits should be corrected with a {\it varying} hardening factor $1.7<f_{\rm col}<3$ in order to recover the actual value of the inner disc radius.

It should be clear, however, that the issue of determining the exact shape of accretion disc 
spectra is very complicated and is very sensitive to the actual physical models used to 
describe it. In particular, key elements are the vertical structure of the disc, its density 
profile and, in particular, the surface density.   

It is quite possible, and perhaps easier to conceive, that dramatic changes in the disc spectrum  are produced by changes in the surface properties of the accretion disc rather than by the disappearance of the entire inner regions. This, unfortunately, is hard to 
model theoretically. It would be interesting, for example, to assess the problem in the 
extreme case when almost all the accretion power is dissipated in the corona (i.e.\ in the 
limit $f \rightarrow 1$). As we believe that even in our most extreme simulation (S11) with $f=0.8$ the simple treatment of the vertical disc structure we used here can be inadequate, and as this must be true {\it a fortiori} for larger values of $f$, we do not push our study further.

To do that we would need to take into account the X-ray reflection 
spectra of discs illuminated by a hot corona and to model carefully the heating of the disc 
atmosphere (e.g., see Sincell \& Krolik \shortcite{SK97}, Ross \& Fabian \shortcite{RF93}, who show how a hard tail in the disc 
spectrum is produced in this case). It is natural to believe, given the trend we observe in 
the simulations presented above and the observational results we referred to, that in such a situation the oversimplified multicolour 
model will give even smaller values of the colour radius.  

A detailed treatment of this kind requires further numerical work which 
is beyond the aim of this Letter.

\section*{Acknowledgments}
This work was done in the research network
``Accretion onto black holes, compact stars and protostars"
funded by the European Commission under contract number 
ERBFMRX-CT98-0195'. AM, RRR and ACF thank the PPARC, the College of the Holy Cross and the Royal Society 
for support, respectively.

\bsp

\label{lastpage}

\end{document}